\documentclass[12pt]{article}
\usepackage{amstext}
\usepackage{amssymb}
\usepackage{amsmath}
\usepackage{latexsym}

\newtheorem{defi}{Definition}[section]
\newtheorem{lem}[defi]{Lemma}
\newtheorem{prop}[defi]{Proposition}
\newtheorem{theorem}[defi]{Theorem}
\newtheorem{cor}[defi]{Corollary}

\newcommand{\bedefin}{\begin{defi}}
\newcommand{\findefi}{\end{defi} \medskip}

\newcommand{\betheo}{\begin{theorem}$\!\!${\bf \,\,\,}}
\newcommand{\entheo}{\end{theorem}}
\newcommand{\enth}{\end{theorem}}

\newcommand{\becor}{\begin{cor}$\!\!${\bf .}}
\newcommand{\encor}{\end{cor}}

\newcommand{\belem}{\begin{lem}$\!\!${\bf .}}
\newcommand{\enlem}{\end{lem}}

\newcommand{\beprop}{\begin{prop}$\!\!${\bf .}}
\newcommand{\enprop}{\end{prop}}

\newcommand{\prf}{\noindent{\bf{\small Proof.}\,\,\,\,}}
\newcommand{\qed}{\hfill $\blacksquare$}

\newcommand{\be}{\begin{equation}}
\newcommand{\en}{\end{equation}}
\newcommand{\ee}{\end{equation}}

\newcommand{\bea}{\begin{eqnarray}}
\newcommand{\ena}{\end{eqnarray}}

\newcommand{\beano}{\begin{eqnarray*}}
\newcommand{\enano}{\end{eqnarray*}}

\newcommand{\bee}{\begin{enumerate}}
\newcommand{\ene}{\end{enumerate}}

\newcommand{\bei}{\begin{itemize}}
\newcommand{\eni}{\end{itemize}}

\newcommand{\betab}{\begin{tabular}}
\newcommand{\entab}{\end{tabular}}

\newcommand{\bd}{\begin{displaymath}}

\newcommand{\h}{{\mathfrak H}}
\newcommand{\kk}{{\mathfrak K}}

\newcommand{\bPsi}{\mbox{\boldmath $\Psi$}}

\newcommand{\btheta}{\mbox{\boldmath $\theta$}}

\newcommand{\bK}{\mathbf K}

\newcommand{\bv}{\mathbf v}

\newcommand{\bF}{\mathbf F}

\newcommand{\R}{\mathbb R}

\newcommand\spr[1]{\langle#1\rangle}
\newcommand\into{\int_{\mathcal O_n}}

\newcommand\Tr{\operatorname{Tr}}

\newcommand\NN{\Cal N}

 \let\Cal\mathcal

\def\H{\relax\ifmmode {\mathcal H}\else${\mathcal H}$\fi}

\newcommand{\Leb}[2]{
  \ifnum #2=1
  L^#1(\R,\,\ud x)
  \fi
  \ifnum #2=2
  L^#1(\R^2,\,\ud^2 \vec{x}\,)
  \fi}

\newcommand{\ud}{\mathrm{d}}

\begin{document}
\baselineskip 18pt plus 2pt minus 2pt

\begin{center}
{\LARGE \bf  Vector Coherent States from\\[3mm]
       Plancherel's Theorem, Clifford \\[5mm]
      Algebras and Matrix Domains}\\[10mm]
\sc  S. Twareque Ali\\[1mm]
{\small\it Department of Mathematics and Statistics\\[1mm]
Concordia University, Montr\'eal\\[1mm]
Qu\'ebec, Canada H4B 1R6}\\[1mm]
{\scriptsize stali@mathstat.concordia.ca}\\[3mm]
\sc Miroslav Engli\v s\\[1mm]
{\small\it M\'U AV \v CR, \v Zitn\'a 25\\[1mm]
11567 Praha 1, Czech Republic}\\[1mm]
{\scriptsize englis{@}math.cas.cz}\\[2mm]
 and\\[2mm]
\sc Jean-Pierre Gazeau\\[1mm]
{\small\it Astroparticules et Cosmologie and LPTMC \\[1mm]
Boite 7020, Universit\'e Paris 7 Denis Diderot\\[1mm]
 F-75251 Paris Cedex 05, France }\\[1mm]
 {\scriptsize gazeau@ccr.jussieu.fr}\\

\end{center}
\thispagestyle{empty}
\addtocounter{page}{-1}

%--------------------------------------------------------------

\vskip 25pt

\begin{abstract}
As a substantial generalization of the technique for constructing canonical and the
related nonlinear and q-deformed coherent states, we present here a method for constructing vector
coherent states in the same spirit. These vector coherent states may have a finite or
an infinite number of components.  As examples we first apply the technique to construct vector
coherent states using the Plancherel isometry for groups
and vector coherent states associated to Clifford algebras, in particular quaternions.
As physical examples, we discuss vector coherent states for a quantum optical model and finally
apply the general technique to build vector coherent states over  certain matrix domains.

\end{abstract}

\newpage

%---------------------------------------------------------------
\baselineskip 18pt plus 2pt minus 2pt
\parskip=6pt

\section{Introduction}\label{sec:intro}
\setcounter{equation}{0}

   The well-known canonical coherent states are defined as (see, for example \cite{aag_book,klauskag,perel}):
\be
  \vert z \rangle = e^{-\frac {\vert z\vert^2}2}\; \sum_{k=0}^\infty \frac {z^k}{[k!]^{\frac 12}}\; \phi_k\; ,
\label{CCS}
\en
where the $\phi_k \;, k = 0 , 1, 2, \ldots , \infty$, form an orthonormal basis in a (complex,
separable, infinite dimensional) Hilbert space $\h$. The related {\em deformed} or {\em nonlinear} coherent states are
the generalized versions:
\be
  \vert z \rangle = {\mathcal N}(\vert z\vert^2)^{-\frac 12}\; \sum_{k=0}^\infty
     \frac {z^k}{[x_k !]^{\frac 12}}\; \phi_k\; ,
\label{DCS}
\en
where the generalized factorial $x_k !$ is the quantity,
$ x_k ! =  x_1 x_2 \ldots x_n$, for a sequence of positive numbers, $x_1 , x_2 , x_3 , \ldots ,$ and by convention,
$x_0 ! =1$. The normalization factor ${\mathcal N}(\vert z\vert^2)$ is chosen so that
$\langle z \vert z \rangle = 1$. The coherent states form an {\em overcomplete} set of vectors in the
Hilbert space $\h$;  there is also the associated resolution of the identity,
\be
 \int_{\mathcal D}d\nu (z, \overline{z} )\; {\mathcal N}(\vert z\vert^2) \; \vert z \rangle\langle z \vert = I\; ,
 \label{resolid}
\en
where $I$ denotes the identity operator on the Hilbert space $\h$, $\mathcal D$ is a convenient domain of the
complex plane (usually the open unit disc, but which could also be the entire plane). The measure $d\nu$ is usually
of the type $d\theta\;d\lambda (r)$ (for $z = re^{i\theta}$), where $d\lambda$
is related to the $x_k !$ through a moment condition (see, for example, \cite{sim97} for an exhaustive discussion
of the moment problem):
\be
  \frac {x_k !}{2\pi} = \int_0^L d\lambda (r )\; r^{2k}\; , \qquad
  \frac 1{2\pi} = \int_0^L d\lambda (r )\; ,
\label{momprob}
\en
$L$ being the radius of convergence of the series $\sum_{k=0}^\infty\frac {z^k}{\sqrt{x_k !}}$. This
means that once the quantities $\rho (k) = x_k !$ are specified, the measure $d\lambda$ is to be determined
by solving the moment problem (\ref{momprob}). An extensive literature exists on the construction of entire families of
coherent states of this type; as a small sampling, we might suggest \cite{gazklau,klaupensix,manmarsuza,odz98}.

   Quite generally, one can start with a function $f(z)$, holomorphic in the open disc
${\mathcal D} = \vert z\vert < L$, and having a Taylor expansion of the type,
\be
  f(z) = \sum_{k=0}^\infty \frac{z^{2k}}{\rho (k)}, \qquad \rho (k) > 0,\;\; \forall\; k\;,\quad
          \rho (0) = 1\; ,
\label{eq:holomorph-fcn}
\en
where the sequence $\{\rho (k)\}_{k=0}^\infty$  satisfies,
\be
  \lim_{k \rightarrow \infty}\;\frac {\rho (k+1)}{\rho (k)} = L^2 > 0\; .
\label{eq:rad-conv}
\en
Then, writing $x_k = \frac {\rho (k)}{\rho (k-1)}$, for $k \geq 1$, and $x_0 = 0$, the  vectors
\be
  \vert z \rangle = f(\vert z\vert)^{-\frac 12}\;\sum_{k=0}^\infty \frac {z^k}{[x_k !]^{\frac 12}}\; \phi_k\; ,
\label{eq:holomorphCS}
\en
define a set of deformed or nonlinear coherent states for all $z \in {\mathcal D}$ which are not zeroes of
$f(z)$. The moment problem (\ref{momprob}) is used to determine the measure
$d\lambda$ and then one has the resolution of the identity,
\be
  \int_0^{2\pi}\!\! d\theta\int_0^L\!\! d\lambda (r)\; f(\vert z\vert )  \;
    \vert z \rangle\langle z \vert = I\; ,
 \label{eq:hol-resolid}
\en
and normalization $\langle z  \vert  z\rangle = 1$.

   It is also known \cite{borzov01,bordam,odz01} that if the sum $\sum_{k=0}^\infty \frac 1{\sqrt{x_k }}$
diverges then the above
family of coherent states is naturally associated to a set of polynomials $\{p_k (x)\}_{k=0}^\infty$, orthogonal
with respect to some measure $dw (x)$ on the real line, which may then be used to replace the $\phi_k$ in the
definition (\ref{eq:holomorphCS}) of the CS. To see this, define the generalized annihilation operator $a_f$
by its action on the vectors $\vert z\rangle$,
\be
  a_f\vert z \rangle = z\vert z \rangle\; ,
\label{eq:gen-anhih-op}
\en
and its adjoint $a_f^\dagger$. Their actions on the basis vectors are easily seen to be:
\be
  a_f\;\phi_k = \sqrt{x_k}\;\phi_{k-1}\; , \qquad a_f^\dagger\; \phi_k = \sqrt{x_{k+1}}\;\phi_{k+1}\; .
\label{eq:ann-creat-op}
\en
Using these we define the operators,
\be
  Q_f = \frac 1{\sqrt{2}}\; [a_f + a_f^\dagger ]\; , \qquad
  P_f = \frac 1{i\sqrt{2}}\; [a_f - a_f^\dagger ]\; ,
\label{eq:pos-mom-op}
\en
which are the deformed analogues of the standard position and momentum operators. The operator $Q_f$ has the
following action on the basis vectors:
\be
  Q_f \phi_k = \sqrt{\frac{x_k}2}\; \phi_{k-1} + \sqrt{\frac{x_{k+1}}2}\; \phi_{k+1}\; .
\label{eq:pos-op-act}
\en
If now the sum $\sum_{k=0}^\infty \frac 1{\sqrt{x_k }}$ diverges, the operator $Q_f$ is essentially
self-adjoint and hence has a unique self-adjoint extension, which we again denote by $Q_f$. Let $E_x , \;
 x\in \mathbb R$, be the spectral family of $Q_f$, so that,
$$ Q_f = \int_{-\infty}^\infty x \; dE_x \; .$$
Thus there is a measure $dw(x)$ on $\mathbb R$ such that on the Hilbert space $L^2 (\mathbb R , dw)$, the action of
$Q_f$ is just a multiplication by $x$. Consequently, on this space, the relation (\ref{eq:pos-op-act})
assumes the form
\be
  x\phi_k (x) = b_k  \phi_{k-1}(x) + b_{k+1} \phi_{k+1} (x)\; , \qquad b_k =  \sqrt{\frac{x_k}2}\;,
\label{eq:pos-op-act2}
\en
which is a two-term recursion relation, familiar from the theory of orthogonal polynomials. It follows that
$dw (x) = d\langle \phi_0\vert E_x \phi_0\rangle $, and the $\phi_k$ may be realized as the polynomials obtained
by orthonormalizing the sequence of monomials $1, x, x^2 , x^2 , \ldots\; , $ with respect to this measure
(using a Gramm-Schmidt procedure). Let us use the notation $p_k (x)$ to write the vectors $\phi_k$, when they are
so realized, as orthogonal polynomials in $L^2 (\mathbb R , dw)$. Then, for any $w$-measurable set
$\Delta \subset \mathbb R$,
\be
  \langle\phi_k \vert E(\Delta)\phi_\ell\rangle = \int_{\Delta} dw(x)\; p_k (x)p_\ell (x)\; ,
\label{eq:poly-basis}
\en
and
\be
    \langle\phi_k \vert \phi_\ell\rangle = \int_{\mathbb R} dw(x)\; p_k (x)p_\ell (x)= \delta_{k\ell}\; .
\label{eq:poly-orthog}
\en
Also setting $\eta_z = \vert z \rangle$,
\be
  \eta_z (x ) = f(\vert z\vert )^{-\frac 12}\;\sum_{k=0}^\infty \frac {z^k}{[x_k !]^{\frac 12}}\; p_k (x) \; ,
\label{eq:orth-polynCS}
\en
and then
\be
  G(z, x) = \sum_{k=0}^\infty \frac {z^k}{[x_k !]^{\frac 12}}\; p_k (x) \; ,
\label{eq:gen-fcn}
\en
is the generating function for the polynomials $p_k$. Note that in the original definition of the CS in
(\ref{eq:holomorphCS}), the vectors $\phi_k$ were simply an arbitrarily chosen orthonormal basis in an
abstract Hilbert space $\h$. As such, we may use any family of orthogonal polynomials to replace them in
(\ref{eq:orth-polynCS}) and then (\ref{eq:gen-fcn}) would give the generating function for this set of polynomials.
However, the set obtained by using the recursion relations (\ref{eq:pos-op-act2}) is in a sense canonically related
to the family of CS $\vert z \rangle$.

   In the present paper we intend to extend many of these considerations to vector coherent states.

\section{A general construction for VCS}\label{sec:gen-cons-vcs}
\setcounter{equation}{0}

  Vector coherent states (VCS) have been studied widely in the literature (see, e.g., \cite{aag_book} for
a general discussion). Generally, these are multicomponent
coherent states, $\vert x , i\rangle$, where $x$ ranges through some continuous parameter space and $i$ is a
discrete index (usually finite). A method for constructing VCS over matrix domains, where essentially, the
variable $z$ in (\ref{DCS}) is replaced by a matrix valued function, has been developed in \cite{thiali}.
We adopt below a more general definition for such states, which will then include
(as special cases) the coherent states of the type mentioned above as well as all other types of coherent
states presently appearing in the literature. In particular, elements from certain interesting matrix domains
will be used in place of $z$ to build $n$-component VCS.

   We will denote our parameter space for defining VCS by $X$ which will be a space with a topology ( usually a
locally compact space), equipped with a measure $\nu$. Let $\h$ and $\kk$ be two (complex, separable) Hilbert spaces,
of infinite or finite dimensions, which we denote by $\text{dim}(\h )$ and  $\text{dim}(\kk )$, respectively.
In $\h$ we specify an orthonormal basis $\{\phi_k\}_{k=0}^{\text{dim}(\h )}$ and in $\kk$ we take an
orthonormal basis $\{\chi^i\}_{i=1}^{\text{dim}(\kk )}$. Let ${\mathcal B}_2 (\kk)$ denote the vector space of all
Hilbert-Schmidt operators on $\kk$. This  is a Hilbert space under the scalar product
$$ \langle Y \vert Z \rangle_2 = \text{Tr}[Y^* Z ]\; , \qquad Y , Z \in {\mathcal B}_2 (\kk)\; ,$$
$\text{Tr}$ denoting the trace,
$$
 \text{Tr} [Z ] =
    \sum_{i=1}^{\text{dim}(\kk)}\langle\chi^i\;\vert \;Z \; \chi^i\rangle\; .
$$
Let $F_k :X \longrightarrow {\mathcal B}_2 (\kk) , \;\; k = 0, 1,2, \ldots \text{dim}(\h )$, be a set of continuous
mappings satisfying the two conditions:
\begin{enumerate}
\item[$(a)$]
for each $x \in X$,
\be
   0 < {\mathcal N}(x) = \sum_{k=0}^{\text{dim}(\h )} \text{Tr} [\vert F_k (x)\vert^2 ] < \infty\; ,
\label{F-cond1}
\en
where $\vert F_k (x)\vert = [F_k (x)F_k (x)^*]^{\frac 12}$ denotes the positive part of the operator $F_k(x)$;

\item[$(b)$]
if $I_\kk$ denotes the identity operator on $\kk$ then,
\be
  \int_X d\nu (x) \; F_k (x)F_\ell (x)^* = \delta_{k\ell}\; I_\kk\; , \qquad k , \ell =0, 1,2, \ldots ,
  \text{dim}(\h )\; ,
\label{F-cond3}
\en
the integral converging in the weak sense.
\end{enumerate}

It is not hard to see that as a consequence of (\ref{F-cond1}),
 for each $x \in X$, the linear map, $T(x): \kk \longrightarrow \kk\otimes\h$, defined by
\be
  T(x)\chi = {\mathcal N}(x)^{-\frac 12}\; \sum_{k=0}^{\text{dim}(\h)}
          F_k (x)\; \chi\otimes \phi_k \; , \qquad \chi \in \kk\; ,
\label{F-cond2}
\en
is bounded.

   Vector coherent states (VCS), $\vert x ; \chi \rangle \in \kk \otimes \h$, are
now defined for each $x \in X$ and $\chi \in \kk$ by the relation,
\be
  \vert x ; \chi \rangle = T(x)\chi = {\mathcal N}(x)^{-\frac 12}\; \sum_{k=0}^{\text{dim}(\h)}
          F_k (x)\; \chi\otimes \phi_k \; .
\label{def-vcs}
\en
In particular, we single out the VCS
\be
  \vert x ; i \rangle := \vert x ; \chi^i \rangle\; , \qquad i = 1,2, \ldots \text{dim}(\kk)\; .
\label{basis-vcs}
\en
For fixed $x \in X$, the $\vert x ; i \rangle$ may not all be linearly independent and some may even
be zero, but any VCS $\vert x ; \chi\rangle$ can always be written as a linear combination,
\be
  \vert x ; \chi \rangle = \sum_{i=1}^{\text{dim}(\kk)}c_i\;\vert x ; i \rangle \; , \quad \text{where} \quad
     \chi =  \sum_{i=1}^{\text{dim}(\kk)}c_i\;\chi^i\; , \quad c_i \in \mathbb C\; .
\label{basexp-vcs}
\en
Moreover, as we shall see below, the set of all VCS, as $x$ runs through $X$ and $i =1,2, \ldots , \text{dim}(\kk)$,
constitutes an {\em overcomplete} family of vectors in $\kk\otimes\h$. Indeed, we have immediately the result,
\betheo
  The VCS $\vert x ; i \rangle$ satisfy the
\begin{enumerate}
\item[$(a)$] normalization condition,
\be
   \sum_{i=1}^{\text{dim}(\kk)}\Vert\; \vert x ; i \rangle \;\Vert^2 = 1\; ,
\label{norm-vcs}
\en
and
\item[$(b)$]
resolution of the identity,
\be
    \sum_{i=1}^{\text{dim}(\kk)}\int_X d\nu (x)\;  {\mathcal N}(x)\;\vert x ; i \rangle\langle x ; i \vert =
           I_\kk \otimes I_\h\; ,
\label{resolid-vcs}
\en
the sum and the integral converging in the weak sense.
\end{enumerate}
\entheo
\prf
  The proof is absolutely straightforward, however a quick demonstration is still in order. For part (a),
\beano
  \sum_{i=1}^{\text{dim}(\kk)}\Vert\; \vert x ; i \rangle \;\Vert^2 & = &
    \sum_{i=1}^{\text{dim}(\kk)}\langle x ; i\; \vert\; x ; i \rangle \\
    & = & {\mathcal N}(x)^{-1}\;\sum_{i=1}^{\text{dim}(\kk)}\;\sum_{k,\ell = 0}^{\text{dim}(\h)}
         \langle\chi^i\vert F_k (x)^* F_{\ell}(x)\chi^i \rangle\;\langle\phi_k\vert\phi_\ell\rangle \\
    & = & {\mathcal N}(x)^{-1}\;\sum_{i=1}^{\text{dim}(\kk)}\;\sum_{k=0}^{\text{dim}(\h)}
          \langle\chi^i\vert F_k (x)^* F_k (x)\chi^i \rangle\; .
\enano
Since all the terms within the summations are positive, the two sums may be interchanged and then using
(\ref{F-cond1}) we immidiately get (\ref{norm-vcs}). To prove part (b), let $A$ denote the
formal operator represented by the sum and integral on the left hand side of (\ref{resolid-vcs}). Let $\chi , \xi \in
\kk$ and $\phi , \psi \in \h$ be arbitrary. Then, from the definition of weak convergence we have,
\beano
\langle\chi\otimes\phi\;\vert\; A(\xi\otimes\psi)\rangle & = &
   \sum_{i=1}^{\text{dim}(\kk)}\int_X d\nu (x)\; {\mathcal N}(x)\langle\chi\otimes\phi\; \vert\; x;i\rangle
      \langle x; i\; \vert\; \xi\otimes\psi\rangle \\
   & = & \sum_{i=1}^{\text{dim}(\kk)}\int_X d\nu (x)\; \Big[\sum_{k,\ell =0}^{\text{dim}(\h)}
   \langle\chi\vert F_k (x)\chi^i\rangle\;
           \langle\chi^i \vert F_\ell (x)^* \xi\rangle\;\\
        &\qquad&
           \times\;\langle\phi\vert\phi_k\rangle\langle\phi_\ell \vert\psi\rangle\Big]\; .
\enano
The boundedness of the operator $T(x)$ in (\ref{def-vcs}) and the fact that $\sum_{i=1}^{\text{dim}(\kk)}
\vert\chi^i\rangle\langle\chi^i\vert = I_\kk$, allows us to interchange the sum over $i$ with the
integral and the two sums over $k$ and $\ell$. Thus,
$$ \langle\chi\otimes\phi\;\vert\; A(\xi\otimes\psi)\rangle = \int_X d\nu (x)\; \Big[\sum_{k,\ell =0}^{\text{dim}(\h)}
         \langle\chi\vert F_k (x) F_\ell (x)^* \xi\rangle\;
         \langle\phi\vert\phi_k\rangle\langle\phi_\ell \vert\psi\rangle\Big]\; . $$
Again, in view of the boundedness of $T(x)$, the integral and the two summations in the above
expression can be interchanged. Next, taking account of (\ref{F-cond3}) and the relation $\sum_{k=0}^{\text{dim}(\h)}
\vert\phi_k \rangle\langle\phi_k\vert = I_\h$ we obtain
$$ \langle\chi\otimes\phi\;\vert\; A(\xi\otimes\psi)\rangle = \langle\chi\vert\xi\rangle\;
     \langle\phi\vert\psi\rangle\; , $$
proving (\ref{resolid-vcs}).
\qed

   There is a reproducing kernel, $K: X \times X \longrightarrow \mathcal L (\mathfrak K)$ (bounded operators on
$\mathfrak K$), naturally associated with the family of VCS (\ref{def-vcs}). It is given by
\be
   K(x, y) =  \sum_{k=0}^\infty F_k (x)^* F_k (y)\; .
\label{matrepker}
\end{equation}
Note that for each $(x, y)\; , K(x, y)$ is a bounded operator on $\mathfrak K$. It has the properties
\bea
K(x,y)^* & = & K(y,x)^*\; , \label{repkerprop1}\\
\int_X d\nu (y)\; K(x, y)K(y, z) & = & K(x,z)\; ,
\label{repkerprop2}
\ena
the integral relation  (\ref{repkerprop2}) following immediately from  (\ref{F-cond3})
and (\ref{resolid-vcs}). If in addition, the kernel satisfies
\be
   \langle \chi \; \vert\; K(x, x )\chi\rangle > 0 \;, \quad \forall \chi \neq 0\;,
\label{kerpos}
\end{equation}
that is, $K(x,x)$ is a strictly positive operator, then the vectors (\ref{basis-vcs})
are linearly independent, for each fixed $x \in X$.

\section{Example based on the Plancherel isometry}\label{sec:ex-planch-isom}
\setcounter{equation}{0}

Suppose that $G$ is a locally compact group with Type-I regular representation. Let $U(g), \;\; g \in G$ be a
subrepresentation of the left regular representation, acting on the Hilbert space $\kk$. Assume $U(g)$ to be
multiplicity free, such that it has the decomposition into irreducibles
\be
   U(g) = \int_{\widehat{\Sigma}}^\oplus d\nu_G (\sigma )\;U_\sigma (g)\; , \qquad
   \kk = \int_{\widehat{\Sigma}}^\oplus d\nu_G (\sigma )\;\kk_\sigma\; ,
\label{U-decomp}
\en
where $\nu_G$ is the Plancherel measure on the dual $\widehat{G}$ of the group and
$\nu_G (\widehat{\Sigma}) < \infty$. The irreducible representations $U_\sigma (g)$ are carried by the
Hilbert spaces $\kk_\sigma$; the measure $\nu_G$ could have a discrete part so that the integrals in
(\ref{U-decomp}) could also include sums. There exists \cite{DuMo,Tat} on
($\nu_G$-almost) all $\kk_\sigma$, a positive,
self-adjoint operator $C_\sigma$, called the Duflo-Moore operator with the property that if $G$ is unimodular
then $C_\sigma$ is a multiple of the identity operator on $\kk_\sigma$, while if $G$ is non-unimodular then
it is a densely defined unbounded operator with densely defined inverse. Set
\be
  C = \int_{\widehat{\Sigma}}^\oplus d\nu_G (\sigma )\;C_\sigma\; ,
\label{duflo-moore-op}
\en
and let $\text{Dom}(C)$ denote its domain. Any vector $\chi \in \kk$ has components $\chi_\sigma \in \kk_\sigma$
and
 $$  \Vert\chi\Vert^2 = \int_{\widehat{\Sigma}}^\oplus d\nu_G (\sigma )\;\Vert\chi_\sigma\Vert_\sigma^2\; , $$
$\Vert\ldots\Vert_\sigma$ denoting the norm in $\kk_\sigma$. Then, as a
consequence of Plancherel's theorem, for all $\eta , \eta' \in \text{Dom}(C)$ and $\chi, \chi'\in \kk$,
the following orthogonality relation holds \cite{AFK-2001}:
\be
 \int_G d\mu(g)\; \overline{\langle U(g)\eta'\vert\chi'\rangle}\;\langle U(g)\eta\vert\chi\rangle
       = \int_{\widehat{\Sigma}} d\nu_G (\sigma)\; \langle C_\sigma\eta_\sigma\vert C_\sigma\eta'_\sigma\rangle\;
               \langle\chi'_\sigma\vert \chi_\sigma\rangle \; ,
\label{planch-orthog}
\en
where $d\mu$ denotes the (left invariant) Haar measure of $G$. Thus, if we choose $\eta = \eta'$ and satisfying
$\Vert C_\sigma\eta_\sigma\Vert^2 =1$, for almost all $\sigma \in \widehat{\Sigma}$ (w.r.t. the Plancherel measure
$\nu_G$), then we obtain the resolution of the identity,
\be
  \int_G d\mu (g)\; \vert U(g)\eta\rangle\langle U(g)\eta\vert = I_\kk \; .
\label{planch-resolid}
\en
(Note, if $G$ is non-unimodular, each $C_\sigma$ is an unbounded operator, and the condition
$\nu_G (\widehat{\Sigma}) < \infty$ could be relaxed
\cite{Fu}. If however $G$ is unimodular, each $C_\sigma$ is a
multiple of the identity and the condition $\nu_G (\widehat{\Sigma}) < \infty$ becomes necessary).

  Let $\eta^k \in \kk , \; \; k = 0,1, 2, \ldots , \text{dim}(\h)$, be
{\em mutually orthogonal vectors\/,} chosen so that
\begin{enumerate}
\item[$(1)$] for each $k,\;\;  \eta^k \in \text{Dom}(C)\; ,$
\item[$(2)$] for each $k$ and almost all $\sigma \in \widehat{\Sigma},\;\; \Vert C_\sigma\eta^k_\sigma\Vert^2 =1$.
\end{enumerate}
Define
\be
   V_k (g) = \frac 1{\Vert\eta^k\Vert}U(g)\vert\eta^k\rangle\langle\eta^k\vert \in {\mathcal B}_2 (\kk )\; .
\label{v-func1}
\en
Then,
\be
   \int_G d\mu(g)\;V_k (g) V_\ell (g)^* = \delta_{k\ell}\;I_\kk \qquad \text{and}\qquad
     \text{Tr}[V_k (g)V_k (g)^*] = \Vert\eta^k\Vert^2\; ,
\label{v-func2}
\en
the first relation following from the orthogonality of the $\eta^k$ and (\ref{planch-resolid}).
Note that $\h$ is in general an abstract Hilbert space, different from $\kk$; however, its dimension
cannot exceed that of $\kk$. Let us choose an orthonormal basis, $\{\chi^i\}_{i=1}^{\text{dim}(\kk)}$ in
$\kk$, not necessarily related to  the vectors $\{\eta^k\}$ and a second orthonormal basis,
$\{\phi_k\}_{k =0}^\infty$ in $\h$. In order to construct VCS, it is generally necessary to add a second
locally compact space $R$, equipped with a (Radon) measure $\lambda$, to the group $G$. Let $f_k , \;\; k = 0,1,2,
\ldots, \text{dim}(\h)$, be a sequence of continuous complex functions in the Hilbert space $L^2 (R, d\lambda )$
satisfying,
\begin{enumerate}
\item[$(1)$] for all $k,\;\;  \Vert f_k\Vert^2 =1\; ;$
\item[$(2)$] for each $r$ in the support of the measure $\lambda$,
\be
   0 \neq \sum_{k=0}^{\text{dim}(\h )} \vert f_k (r)\vert^2\; \Vert\eta^k\Vert^2 < \infty\; .
\label{func-normaliz}
\en
\end{enumerate}
Let $X = R\times G$ and $\nu = \lambda \otimes\mu$. Then, writing $x = (r,g )$ and $F_k (x) =
f_k (r)\;V_k (g)$, the set
\be
  \vert x; i\rangle = {\mathcal N}(r)^{-\frac 12}\;\sum_{k=0}^{\text{dim}(\h )}F_k (x)\chi^i\otimes \phi_k \; ,
  \qquad x \in X, \;\; i=1,2, \ldots , \text{dim}(\kk ) \; ,
\label{planch-vcs}
\en
with
\be
  {\mathcal N}(r) = \sum_{k=0}^{\text{dim}(\h )}\vert f_k (r)\vert^2\; \Vert\eta^k\Vert^2\; ,
\label{planch-vcs-norm}
\en
is easily seen to define a family of VCS.

  Note that taking $\h$ to be a one-dimensional space, the above type of VCS can be used to derive the usual
Gilmore-Perelomov CS or the sort of VCS discussed in \cite{aag_book}.

  As an explicit example, we construct a family of VCS of the above type using the principal series
representations of $G= SU(1,1)\; (\simeq SL(2, \mathbb R ))$. This group is unimodular; an element $g \in
SU(2,2)$ has the form,
$$
  g = \begin{pmatrix} \alpha & \beta \\ \overline{\beta} & \overline{\alpha}\end{pmatrix}\; , \qquad
   \alpha , \beta \in \mathbb C\; , \quad \vert\alpha\vert^2 - \vert\beta\vert^2 = 1\; .
$$
In terms of the parametrization,
$$ g = r(\phi)\; a(t)\; r (\psi )\; , \qquad 0\leq \phi \leq 2\pi\; , \quad -2\pi \leq \psi
< 2\pi\; , \quad t \in \mathbb R\; , $$ where,
$$
r(\varphi ) = \begin{pmatrix} e^{\frac {i\varphi}2} & 0 \\ 0 & e^{-\frac {i\varphi}2}\end{pmatrix}\; ,
\qquad a(t) = \begin{pmatrix} \cosh\frac t2 & \sinh\frac t2 \\ \sinh\frac t2 & \cosh\frac t2 \end{pmatrix}\; , $$
the Haar measure is $d\mu = \sinh t\; dt\;d\phi\;d\psi$. Denote by $U_\text{reg}$ the regular representation
of this group on $L^2 (G, d\mu )$:
$$ (U_\text{reg} (g)f)(g') = f(g^{-1}g' )\; , \qquad f \in L^2 (G, d\mu )\; . $$
For any unitary irreducible representation  $U_\sigma$ of $SU(1,1)$, acting on the Hilbert space
$\h_\sigma$, the operator
$$ U_\sigma (f) = \int_G d\mu (g) f(g) U_\sigma (g)\; , \qquad f \in L^1 (G, d\mu )\cap L^2 (G, d\mu )\; ,$$
is Hilbert-Schmidt and the Plancherel formula (see, for example, \cite{lipsman}) may be written as,
\bea
  \int_G\; d\mu (g) \vert f(g)\vert^2 & = & \frac 1{4\pi^2}\;\bigg[\int_0^\infty \sigma \tanh\pi\sigma\;d\sigma\;
  \Vert U_\sigma^{(0)} (f)\Vert^2_2 \nonumber\\
       & \quad  + & \int_0^\infty \sigma \coth\pi\sigma\;d\sigma\;
  \Vert U_\sigma^{(\frac 12)} (f)\Vert^2_2 \:\bigg]\nonumber\\
        & \quad  + & \sum_{n\geq 1,\; n \in \frac 12 \mathbb Z}\frac {2n - 1}{8\pi^2}\;
  \left[\Vert U^+_n (f)\Vert^2_2 + \Vert U^-_n (f)\Vert^2_2\: \right]\; ,
\label{su11-planch}
\ena
$\Vert \ldots \Vert_2$ denoting the Hilbert-Schmidt norm.
In this formula, which essentially expresses the decomposition of $U_\text{reg}$ into irreducibles,
the continuously labelled representations, $U^{(\varepsilon )}_\sigma , \;\; \varepsilon = 0, \frac 12 ,  \;\;
\sigma \in \mathbb R^+$, are elements of the principal series, while the discretely labelled
$U^\pm_n$ are (almost all) elements of the discrete series (the `+' corresponding to the holomorphic and
the `-' to the anti-holomorphic representations). The complementary series of representations constitute
a set of Plancherel measure zero and hence do not appear in the above decomposition. (This is a general
feature of the theory of representations of non-compact semi-simple Lie groups.)

  The principal series representations $U^{(\varepsilon )}_\sigma$ are all carried by the Hilbert space
$\h^{(\varepsilon )}_\sigma \simeq L^2 (S^1 , d\theta / 2\pi )$, acting in the manner,
\be
   (U^{(\varepsilon )}_\sigma (g)\psi )(e^{i\theta}) =
   \left[-\overline{\beta}e^{i\theta} + \alpha \right]^{\varepsilon -\frac 12 +i\sigma }\;
   \left[-\beta e^{-i\theta} + \overline{\alpha} \right]^{-\varepsilon -\frac 12 +i\sigma }\;
   \psi (g^{-1}e^{i\theta})\; ,
\label{suii-prin-ser}
\end{equation}
where,
$$ g^{-1}e^{i\theta} = \frac {\overline{\alpha}e^{i\theta} - \beta}{-\overline{\beta}e^{i\theta} +\alpha}\; . $$
Let $\widehat{\Sigma} \subset \mathbb R^+$ have finite Plancherel measure, i.e.,
$$ \frac 1{4\pi^2}\:\int_{\widehat{\Sigma}} d\sigma\; \sigma \tanh\pi\sigma < \infty\; ,$$
and consider the corresponding  subrepresentation $U$ of $U_\text{reg}$:
$$ U(g) = \frac 1{4\pi^2}\:\int^{\oplus}_{\widehat{\Sigma}}d\sigma\;
              \sigma \tanh\pi\sigma\; U^{(0)}_\sigma (g)\; . $$
This representation is carried by the Hilbert space $\mathfrak K =
L^2 (\widehat{\Sigma} ,\:\sigma \tanh\pi\sigma\;d\sigma/4\pi^2 )\otimes L^2 (S^1 , \: d\theta / 2\pi )$, with trivial action
on the first space and the action (\ref{suii-prin-ser}) on the second. For the vectors $\eta^k$ we choose the Fourier
orthonormal exponentials, $e^{ik\theta}, \;\; k \in \mathbb Z$:
$$ \eta^k = \mathbb I \otimes \vert e^{ik\theta} \rangle\; , \quad \text{where}\quad
    \mathbb I ( \sigma ) = 1 , \; \; \forall \sigma \in \widehat{\Sigma}\; .$$
Following (\ref{v-func1}), the operators $V_k$ now have the form,
\bea
  V_k (g)  & = & \frac 1{4\pi^2}\:\int_{\widehat{\Sigma}} d\sigma\; \sigma \tanh\pi\sigma \;
            U^{(0)}_\sigma (g)\vert e^{ik\theta} \rangle\langle e^{ik\theta}\vert\nonumber\\
           & = & \frac 1{4\pi^2}\:\int_{\widehat{\Sigma}} d\sigma\; \sigma \tanh\pi\sigma \;
           \vert -\overline{\beta}e^{i\theta} + \alpha\vert^{2i\sigma - 1}\; \vert (g^{-1}e^{i\theta})^k\rangle
           \langle e^{ik\theta}\vert\; ,
\label{suii-prin-ser2}
\ena
(where, with a slight abuse of notation, we have dropped an implicit tensor product). The integration over
$\sigma$ can be performed explicitly:
\begin{align}
& \int_{\widehat{\Sigma}} d\sigma\; \sigma \tanh\pi\sigma \;
           \vert -\overline{\beta}e^{i\theta} + \alpha\vert^{2i\sigma}\nonumber \\
           & = \left\{ e^{\sigma \mathcal Z }\left[
           \frac \sigma{\mathcal Z} + \frac 3{\mathcal Z^2} + 2\sum_{n\geq 1}(-1)^n\left[
           \frac \sigma{\mathcal Z -2n} +  \frac 1{(\mathcal Z -2n )^2}\right] e^{-2n\sigma}
           \right]\right\}_{\partial\widehat{\Sigma}}\; ,\nonumber\\
           & \qquad \qquad \mathcal Z = 2i\ln \;\vert -\overline{\beta} e^{i\theta} + \alpha \vert\; .
\label{suii-prin-ser3}
\end{align}
Next, we choose an arbitrary orthonormal basis $\{\chi^i\}_{i \in \mathbb Z^+}$ in $\mathfrak K$ and a second
orthonormal basis $\{\phi_k\}_{k \in \mathbb Z}$ in $\h$. Furthermore, to avoid divergence of
the normalization factor, we adopt the following choice of the vectors $\{f_k\}_{k \in \mathbb Z}
\subset L^2 (R , d\lambda )$ (see \ref{func-normaliz}): we take  $R = \mathbb R , \;\;
d\lambda (r) = \sqrt{\frac \epsilon\pi}\; e^{-\epsilon \;r^2}\; dr$, where $\epsilon > 0$ is a constant and
\be
  f_k (r) = e^{kr}\; e^{-\frac {k^2}{2\;\epsilon}}\; .
\label{suii-prin-ser5}
\end{equation}
Thus, the  normalization constant $\mathcal N$ satisfies
\be
0 <  \mathcal N (r) = \sum_{k \in \mathbb Z} \vert f_k(r)\vert^2\; \Vert\eta^k\Vert^2 =
               \sum_{k \in \mathbb Z}e^{2kr}\;e^{-\frac {k^2}\epsilon } < \infty \; ,
\label{suii-prin-ser6}
\end{equation}
and is simply related to the {\em theta function} of the third kind.

   Collecting all these, we can finally write down the VCS as,
\be
  \vert x ; i\rangle = \mathcal N (r)^{-\frac 12}\sum_{k \in \mathbb Z} e^{kr}\;e^{-\frac {k^2}{2\;\epsilon}}\;
            V_k (g) \chi^i\otimes\phi_k\; ,
\qquad x = (r, g) \in \mathbb R \times SU(1,1) , \quad i \in \mathbb Z^+,
\label{suii-prin-ser8}
\end{equation}
with $V_k (g)$ given by (\ref{suii-prin-ser2}) and (\ref{suii-prin-ser3}).

\section{Example based on Clifford algebras}\label{sec:ex-cliff-alg}
\setcounter{equation}{0}

   We take the simplest case of a Clifford algebra ${\mathcal C}\ell ({\mathbb R}^d )$, of ${\mathbb R}^d$. This is
the smallest algebra extending ${\mathbb R}^d$ (a concise discussion on Clifford algebras may, for example,
be found in \cite{huse}). We thus have a linear map, ${\mathcal C}: {\mathbb R}^d
\longrightarrow {\mathcal C}\ell ({\mathbb R}^d )$, such that
\be
  {\mathcal C} (\mathbf v )^2 = \Vert \mathbf v \Vert^2\; I_{\mathcal C}\; , \qquad \mathbf v \in
                  {\mathbb R}^d\; ,
\label{cliff-cond}
\en
$I_{\mathcal C}$ denoting the identity in the algebra. Let $e_\alpha , \; \; \alpha = 1, 2, \ldots , d$, be the canonical
basis of ${\mathbb R}^d$, in terms of which  $\mathbf v = \sum_{\alpha =1}^d v^\alpha e_\alpha , \;\;
v^\alpha \in \mathbb R$, and we write ${\mathcal C}(e_\alpha ) = {\mathcal C}_\alpha$. Then it follows from
(\ref{cliff-cond}) that,
\be
  \{{\mathcal C}_\alpha , {\mathcal C}_\beta\} =
   {\mathcal C}_\alpha{\mathcal C}_\beta + {\mathcal C}_\beta{\mathcal C}_\alpha
    = 2\delta_{\alpha \beta}\;I_{\mathcal C}\; ,
\label{cliff-cond2}
\en
and generally,
\be
   \{{\mathcal C}(\mathbf v_1 ) , {\mathcal C}(\mathbf v_2 )\} = 2\mathbf v_1 \cdot \mathbf v_2 \;I_{\mathcal C}\; .
\label{cliff-cond3}
\en
We denote the unit sphere of $\mathbb R^d$ by $S^{d-1}$ and points on it by $\widehat{v} , \;\; \Vert\widehat{v}\Vert
= 1$. Then, ${\mathcal C}(\widehat{v})^2 = I_{\mathcal C}$. Suppose that we have a representation of the
algebra ${\mathcal C}\ell ({\mathbb R}^d )$, by $N\times N$ matrices, ${\mathcal C}(\mathbf v )\longmapsto
H(\mathbf v )$, so that $H(\mathbf v )^2 = \Vert {\mathbf v}\Vert^2\; \mathbb I_N$.
We also assume that the generating matrices
$H_\alpha = H(e_\alpha ) , \;\; \alpha = 1, 2, 3, \ldots , d$,
are Hermitian.

   Identifying ${\mathbb R}^d$ with ${\mathbb R}^+ \times S^{d-1}$, we shall use polar coordinates to parametrize
its points:
\be
   \mathbf v = (r, \btheta , \phi) , \quad r \in {\mathbb R}^+ \;, \quad
    \btheta = (\theta_1 , \theta_2 , \ldots \theta_{d-2} ) \in [0, \pi ]^{d-2}\; , \quad
    \phi \in [0 , 2\pi )\; . \label{polar-coord1}
\en
The connection with the Cartesian coordinates $\mathbf v = (v_1 , v_2 , \ldots , v_d )$ is then given by the well-known
equations,
\bea
v_1 & = & r\sin\theta_{d-2}\sin\theta_{d-3}\ldots \sin\theta_1\cos\phi\; , \nonumber\\
v_2 & = & r\sin\theta_{d-2}\sin\theta_{d-3}\ldots \sin\theta_1\sin\phi\; , \nonumber\\
& & \vdots\nonumber\\
v_i & = & r\sin\theta_{d-2}\ldots \sin\theta_{i-1}\cos\theta_{i-2}\; , \qquad 3\leq i \leq d-1\; ,\nonumber\\
& & \vdots\nonumber\\
v_d & = & r\cos\theta_{d-2}\; .\label{polar-coord2}
\ena
Thus, $\Vert\mathbf v\Vert = r$ and
\be
 \mathbf v = r\cos\theta_{d-2} \;e_d + r \sin\theta_{d-2}\; e(\widehat{n})\; ,
\label{polar-coord3}
\en
where $\widehat{n}$ is the vector in $S^{d-2}$
\be
\widehat{n} = \begin{pmatrix}
  n_1 \\ n_2 \\ \vdots\\ n_i \\ \vdots \\ n_{d-1} \end{pmatrix} =
\begin{pmatrix}
  \sin\theta_{d-3}\ldots \sin\theta_1\cos\phi \\
  \sin\theta_{d-3}\ldots \sin\theta_1\sin\phi \\
  \vdots\\
  \sin\theta_{d-3}\ldots \sin\theta_{i-1}\cos\theta_{i-2}\\
  \vdots\\
  \cos\theta_{d-3}\; , \label{polar-coord4}
\end{pmatrix}
\en
and $e(\widehat{n}) = n_1 e_1 + n_2 e_2 + \ldots + n_{d-1} e_{d-1}$. The $SO(d)$ invariant
measure on $\mathbb R^d$ is $r^{d-1}\;dr\; d\Omega (\btheta , \phi)$, where $d\Omega$ is
the invariant ``surface'' measure on $S^{d-1}$:
\be
  d \Omega (\btheta , \phi)
      = \prod_{i=2}^{d-1}\;\sin^{d-i}\theta_{d-i}\;d\theta_{d-i}\;d\phi\; ,
\label{sphere-sur1}
\en
with total ``surface area'':
\be
   \int_{S^{d-1}}d \Omega (\btheta , \phi) = \frac {2\pi^{\frac d2}}{\Gamma (\frac d2 )}\; .
\label{sphere-sur2}
\en

   Going back now to the construction of VCS, using the Clifford algebra ${\mathcal C}\ell ({\mathbb R}^d )$,
we take $X = S^1 \times {\mathbb R}^d$ and, to each element
$x = (\xi , \mathbf v ) \in X$, we associate the $N\times N$ matrix,
\be
  \mathfrak Z (x) = \mathfrak Z (\xi , \mathbf v ) = r[\cos\xi\; \mathbb I_N + i\sin\xi\; H (\widehat v)]\; , \qquad
  r = \Vert \mathbf v \Vert\; , \qquad \widehat v \in S^{d-1}\; .
\label{cliff-vcs1}
\en
Since $H (\widehat v)^2 = \mathbb I_N$, we get (for any integer $k$),
\be
  \mathfrak Z (\xi , \mathbf v )^k =  r^k\; [\cos (k\xi )\; \mathbb I_N + i\sin (k\xi )\; H (\widehat v)] =
               r^k\; e^{ik\xi H(\widehat v)}\; ,
\label{cliff-vcs2}
\en
and
\be
   \text{Tr}\;[(\mathfrak Z (\xi , \mathbf v )^k)^*\; \mathfrak Z (\xi , \mathbf v )^k ] = Nr^{2k}\; .
\label{cliff-vcs3}
\en
Let $\h$ be a complex (separable) Hilbert space and $\{\phi_k\}_{k=0}^{\text{dim}(\h )}$ an orthonormal
basis of it. Let $\mathfrak K$ denote the ($N$-dimensional) vector space of the representation of the Clifford
algebra ${\mathcal C}\ell ({\mathbb R}^d )$ and let
$\chi^i , \;\; i =1,2, \ldots , N$, be an orthonormal basis of $\mathfrak K$. We fix a sequence of
non-zero, positive numbers, $\{x_k\}_{k=0}^{\text{dim}(\h )}$, with the property that the series
$\sum_{k=0}^{\text{dim}(\h )} \frac {y^k}{\sqrt{x_k !}}\; , \;\;y \in \mathbb R$, converges in some non-empty interval,
$\vert y \vert < L$ and suppose that $d\lambda$ is a measure on $\mathbb R^+$, which satisfies the moment
problem
\be
   \int_0^L d\lambda (r)\;r^{2k} = \frac {\Gamma (\frac d2 )}{4\pi^{\frac {d+2}2}}\; x_k !\; ,
   \qquad k=0, 1, 2, 3, \ldots , {\text{dim}(\h )}.
\label{cliff-vcs4}
\en
Then, defining
\be
  F_k(x) = \frac{2\pi^{\frac {d+2}4}} {[\Gamma (\frac d2 )]^\frac 12}\;\frac {\mathfrak Z (x)^k}{\sqrt{x_k !}}\; ,
\label{cliff-vcs5}
\en
we see that,
\be
 \int_0^{2\pi}\!\!\!\int_0^L\!\!\int_{S^{d-1}}d\xi\; d\lambda (r)\; d\Omega\; F_k (x)\;F_\ell (x)^* =
      \delta_{k\ell}\mathbb I_N\; , \qquad k, \ell =0, 1, 2, 3, \ldots , {\text{dim}(\h )}.
\label{cliff-vcs6}
\en

Thus, we have the result:

\begin{theorem}
The vectors,
\be
  \vert \mathfrak Z (x); i\rangle = {\mathcal N}(r)^{-\frac 12}\; \sum_{k=0}^{{\text dim}(\h )}
       \frac {\mathfrak Z (x)^k}{\sqrt{x_k !}}\;\chi^i\otimes\phi_k \; , \qquad {\mathcal N}(r) =
       \frac {4N\pi^{\frac {d+2}2}} {\Gamma (\frac d2 )}\;\sum_{k=0}^{{\text dim}(\h )}\frac {r^{2k}}{x_k !}\; ,
\label{cliff-vcs7}
\en
$i=1, 2, \ldots , N$, define a set of VCS in ${\mathfrak K}\otimes\h$, for $x = (\xi , r , (\btheta , \phi)) \in
[0, 2\pi )\times [0, L )\times S^{d-1}\;$. These satisfy the resolution of the identity,
\be
    \sum_{i=1}^N\;\int_0^{2\pi}\!\!\!\int_0^L\!\!\int_{S^{d-1}}
    d\xi\; d\lambda (r)\; d\Omega (\btheta , \phi)\;\vert \mathfrak Z (\xi , r , \btheta , \phi); i\rangle
       \langle \mathfrak Z (\xi , r , \btheta , \phi); i\vert =
       \mathbb I_N \otimes I_\h\; .
\label{cliff-vcs8}
\en
\end{theorem}

The particular case of quaternions will be discussed in some detail in the next two sections.

\section{A class of physical examples}
\setcounter{equation}{0}

   The following example is of relevance to the study of the spectra of two-level atomic systems
placed in electromagnetic fields \cite{behuni,daouver} -- the Jaynes-Cummings model in quantum optics
is of this general type. Suppose that $H$ is the Hamiltonian of a two level
atomic system and assume that its eigenvalues constitute two
discrete infinite series of positive numbers (corresponding to the two levels). Assume also that there is no
degeneracy and that the energy eigenvalues are ordered  as follows:
\be
  0 < \varepsilon^i_0 < \varepsilon^i_1 < \varepsilon^i_2 < \ldots \varepsilon^i_k < \ldots \; , \qquad i =1,2 \; .
\label{eq:energy-eval}
\en
Let $\psi_k^i , \;\; i = 1, 2, \;\; k = 0, 1, 2, \ldots , \infty$, be the corresponding eigenvectors, which are
assumed to constitute an orthonormal basis of the Hilbert space $\h_{QM}$ of the quantum system. Let $\h$ be an
abstract, complex (separable), infinite-dimensional Hilbert space and $\{\phi_k\}_{k=0}^\infty$ an orthonormal
basis of it. Consider the Hilbert space $\mathbb C^2\otimes\h$; the set of vectors, $\chi^i\otimes\phi_k\; , \;\;
i=1,2\; ; \;\; k =0, 1,2, \;\ldots\;$, where,
$$ \chi^1 = \begin{pmatrix} 1 \\ 0 \end{pmatrix}\;, \qquad \chi^2 = \begin{pmatrix} 0 \\ 1 \end{pmatrix}\; , $$
forms an orthonormal basis of this Hilbert space. Define the unitary map, $V:\h_{QM} \longrightarrow
\mathbb C^2\otimes\h$, such that, $V\psi_k^i = \chi^i\otimes\phi_k$. Formally, this operator can be written as
\be
   V = \sum_{i=1}^2\;\sum_{k=0}^\infty\vert\chi^i\otimes\phi_k\rangle\langle\psi_k^i\vert\; .
\label{diag-isom}
\en
Writing $H_D = VHV^{-1}$, we see that $H_D$ can be expressed in terms of two self-adjoint operators, $H_1 , H_2$,
on $\h$ in the manner,
\be
  H_D  = \begin{pmatrix} H_1 & 0 \\ 0 & H_2 \end{pmatrix} ,\quad  \text{where} \quad
       H_i\phi_k = \varepsilon^i_k \phi_k \; , \;\; i = 1, 2 \; ; \;\;\; k =0, 1, 2, \; \ldots\; .
\label{diag-hamilt}
\en
Next define the two sets of numbers, $x_k = \varepsilon^1_k - \varepsilon^1_0\; , \;\;
y_k = \varepsilon^2_k - \varepsilon^2_0\; , \;\; k=0, 1,2,\; \dots\; $.
For $z, w$ complex numbers, let $L_1$ be the radius of convergence of the series
$\sum_{k=0}^\infty \frac {z^k}{[x_k !]^{\frac 12}}$ and $L_2$ that of
$\sum_{k=0}^\infty \frac {w^k}{[y_k !]^{\frac 12}}$. Define the domain
$${\mathcal D} = \{ (z, w) \in \mathbb C \times \mathbb C\; \vert\; \vert z\vert < L_1\; , \;\;
              \vert w \vert < L_2\}\;. $$
Let $d\lambda_i\; , \;\; i=1,2$, be two measures on $\mathbb R^+$ which satisfy the moment problems
\be
   \int_0^{L_1} d\lambda_1 (r )\; r^{2k} = \frac {x_k !}{2\pi} \; , \qquad
   \int_0^{L_2} d\lambda_2 (r )\; r^{2k} = \frac {y_k !}{2\pi} \;, \qquad k=0, 1, 2, \; \ldots\;  ,
\label{two-mom-prob}
\en
and with $z = r_1 \;e^{i\theta_1}, \;\; w = r_2 \;e^{i\theta_2}$, define the measure
$d\nu = d\lambda_1 (r_1)\; d\lambda_2 (r_2)\; d\theta_1\; d\theta_2$.
Note that
$$ \int_{\mathcal D} d\nu = 1 \; .$$
Finally define the $2\times 2$ matrices,
\be
R(k) = \begin{pmatrix} x_k ! & 0\\ 0 & y_k ! \end{pmatrix}, \;\; k = 0, 1, 2,\;\ldots \; ,
\qquad \mathfrak Z = \begin{pmatrix} z & 0\\ 0 & w  \end{pmatrix}, \;\; (z, w) \in {\mathcal D}\; .
\label{two-matrices}
\en
Note that the matrices $R(k)$ are positive and invertible. Setting
\be
   F_k (\mathfrak Z ) = R(k)^{-\frac 12}\; {\mathfrak Z}^k\; , \qquad k =0, 1,2, \; \ldots\; ,
\label{phys-F-fcn1}
\en
it is straightforward to verify that
\be
   \int_{\mathcal D}d\nu (\mathfrak Z )\;F_k (\mathfrak Z )F_\ell (\mathfrak Z )^* = \mathbb I_2\; \delta_{k\ell}\; .
\label{phys-F-fcn2}
\en
This leads to the result:
\begin{theorem}
The set of vectors,
\be
  \vert \mathfrak Z ; i \rangle = {\mathcal N}(\mathfrak Z )^{-\frac 12}  \sum_{k=0}^\infty R(k)^{-\frac 12}
             \mathfrak Z^k \chi^i\otimes\phi_k \in {\mathbb C}^2\otimes\h \; ,
\label{qu-opt-vcs}
\en
where,
\be
  {\mathcal N}(\mathfrak Z ) = \sum_{k=0}^\infty \text{Tr}\;
             [F_k (\mathfrak Z )^*\;F_k (\mathfrak Z )] = \sum_{k=0}^\infty \left( \frac {r_1^{2k}}{x_k !}
               + \frac {r_2^{2k}}{y_k !}\right)\; ,
\label{qu-opt-vcs2}
\en
is a family of VCS. They satisfy the resolution of the identity,
\be
  \sum_{i=1}^2\;\int_{\mathcal D} d\nu (\mathfrak Z )\; {\mathcal N}(\mathfrak Z ) \vert \mathfrak Z ; i \rangle
    \langle \mathfrak Z ; i \vert = \mathbb I_2 \otimes I_\h\; .
\label{qu-opt-vcs3}
\en
\end{theorem}

   The above construction can be extended to include a dependence of the coherent states on $SU(2)$ parameters
as well. Indeed, going back to (\ref{phys-F-fcn1}), let $X = {\mathcal D}\times SU(2)$;  denote elements in
$SU(2)$ by $u$ and elements in $X$ by $x = (\mathfrak Z , u )$. Set,
\be
  F_k (x) = uR(k)^{-\frac 12}{\mathfrak Z}^k u^*\; , \qquad x \in X\; .
\label{su2-ext1}
\en
Denote by $d\mu$ the invariant measure on $SU(2)$, normalized to one, and redefine $d\nu$ as
$d\nu (x)  = d\lambda_1 (r_1 )\; d\theta_1\;d\lambda_2 (r_2 )\;d\theta_2\; d\mu (u)$. Then, clearly,
\be
  \int_X d\nu (x)\; F_k (x) F_{\ell} (x)^* = {\mathbb I}_2 \; \delta_{k\ell}\; .
\label{su2-ext2}
\en
Thus, the coherent states,
\bea
  \vert x ; i \rangle & = & {\mathcal N (\mathfrak Z )}^{-\frac 12}\; \sum_{k=0}^\infty
                   F_k (x)\chi^i\otimes\phi_k\nonumber\\
                      & = & {\mathcal N (\mathfrak Z )}^{-\frac 12}\; \sum_{k=0}^\infty
                      u R(k)^{-\frac 12}{\mathfrak Z}^k
                                 u^*\chi^i\otimes\phi_k \in {\mathbb C}^2 \otimes \h \; ,
\label{su2-ext3}
\ena
with $\mathcal N (\mathfrak Z )$ as in (\ref{qu-opt-vcs2}), are well-defined and satisfy the
expected resolution of the identity:
\be
  \sum_{i=1}^2\int_X d\nu (x)\; {\mathcal N (\mathfrak Z )}\vert x ; i \rangle \langle x; i\vert =
                  \mathbb I_2 \otimes I_\h\; .
\label{su2-ext4}
\en

    Finally, it is interesting to replace the matrix $\mathfrak Z$ in (\ref{two-matrices}) by
\be
   \mathfrak Z = u \begin{pmatrix} z & 0 \\ 0 & \overline{z} \end{pmatrix} u^*\;,\qquad u \in SU(2)\; ,
\label{quatern1}
\en
and $R(k)$ by $R(k) = x_k !\; {\mathbb I}_2$. Since a general $SU(2)$ element can be written as
$u = u_{\phi_1}u_\theta u_{\phi_2}$, where,
\be
    u_\theta  = \begin{pmatrix} \cos{\frac \theta 2} & i\sin{\frac \theta 2 } \\
                            i\sin{\frac \theta 2} & \cos{\frac \theta 2} \end{pmatrix}\; , \quad
    u_{\phi_i} = \begin{pmatrix} e^{i\frac {\phi_i}2} & 0 \\ 0 & e^{-i\frac {\phi_i}2} \end{pmatrix}
        \; ,\;\; 0 < \phi_i \leq 2\pi \; , \;\;
                                               0\leq \theta \leq \pi \; ,
\label{quatern2}
\en
we easily get,
\be
    \mathfrak Z = {\mathfrak Z}(z,\overline{z}, \widehat{n} ) = r[\cos\xi \;
               \mathbb I_2 + i\sin\xi\; \sigma (\widehat{n})]\; ,
\label{quatern3}
\en
where we have written
\be
  z = re^{i\xi} \; ,\quad \widehat{n} =
  \begin{pmatrix} \sin\theta \cos\phi \\ \sin\theta \sin\phi \\ \cos\theta \end{pmatrix}\; ,
   \quad \sigma (\widehat{n}) = \begin{pmatrix} \cos\theta & e^{i\phi}\sin\theta  \\
                           e^{-i\phi}\sin\theta & - \cos\theta \end{pmatrix}\; , \;\; \phi = \phi_1\;  .
\label{quatern4}
\en
The associated coherent states are,
\be
 \vert {\mathfrak Z}(z,\overline{z}, \widehat{n} ); i\rangle = {\mathcal N}(r)^{-\frac 12}\sum_{k=0}^\infty
        \frac {{\mathfrak Z}(z, \overline{z}, \widehat{n} )^k}{\sqrt{x_k !}}\chi^i \otimes \phi_k \; , \qquad
       {\mathcal N}(r) = 2\sum_{k=0}^\infty \frac {r^{2k}}{x_k !}\; ,
\label{quatern5}
\en
with the resolution of the identity,
\be
   \frac 1{4\pi}\sum_{i=1}^2 \int_0^L \!\! d\lambda(r)\int_0^{2\pi}\!\! d\xi \int_0^{2\pi} \!\! d\phi \!\int_0^\pi
    \!\! \sin\theta\; d\theta\;
   {\mathcal N}(r) \vert {\mathfrak Z}(z, \overline{z},\widehat{n} ); i\rangle
          \langle {\mathfrak Z}(z, \overline{z}, \widehat{n} ); i \vert = \mathbb I_2 \otimes I_\h\; ,
\label{quatern6}
\en
the measure $d\lambda$, the radius of
convergence $L$ and the $x_k !$ being related by the moment
problem in (\ref{momprob}). If $\xi$ is restricted to $[0, \pi )$,
the resulting set of matrices $\mathfrak Z (z, \widehat{n} )$
yield the $2 \times 2$ complex realization of the quaternions. Consequently, for $x_k = k$ the coherent
states defined in (\ref{quatern5}) are just the  {\em canonical quaternionic coherent states\/} obtained
in \cite{thiali}. We shall generally refer to the vectors (\ref{quatern5}) as {\em quaternionic coherent
states\/}.

\section{Some analyticity properties}\label{sec-analyt-prop}
\setcounter{equation}{0}

   It is well known that the resolution of the identity in (\ref{resolid}) enables one to map the Hilbert
space $\h$, of the coherent states $\vert z \rangle$, unitarily to a Hilbert space of functions which are
analytic in the variable $\overline{z}$. This is done via the mapping $W: \h \longrightarrow
L^2_{\text{a-hol}}({\mathcal D}, d\nu )$,
\be
  (W\phi )(\overline{z}) = {\mathcal N}(r)^{\frac 12}\langle z \vert \phi \rangle =
  \sum_{k=0}^\infty c_k \overline{z}^k\; ,
  \qquad c_k = \frac {\langle \phi_k \vert \phi \rangle}{[x_k !]^{\frac 12}}\; ,
\label{holmap1}
\en
where $L^2_{\text{a-hol}}({\mathcal D}, d\nu )$ is the Hilbert space of all functions holomorphic in
$\overline{z}$ and sqare-integrable with respect to the measure $d\nu$.
The basis vectors $\phi_k$ are mapped in this manner to the monomials $\overline{z}^k /[x_k !]^{\frac 12}$,
and the state $\vert z \rangle$ itself to the function,
\be
  K(\overline{z}', z) = [{\mathcal N}(r'){\mathcal N}(r)]^{\frac 12}\;\langle z' \vert z\rangle =
         \sum_{k=0}^\infty \frac {(\overline{z}' z)^{ k}}{x_k !}\; ,
\label{repker1}
\en
in the variable $\overline{z}'$. Moreover, considered as a function of the two variables $z$ and
$\overline{z}'$,  $K(\overline{z}', z)$ is a reproducing kernel (the analogue of (\ref{matrepker})), satisfying
\be
   \int_{\mathcal D} d\nu (z', \overline{z}')\; K (\overline{z} , z')K(\overline{z}' , z'')
          = K(\overline{z}, z'')\; .
\label{repker2}
\en

    It is interesting to perform a similar transformation for the quaternionic coherent states in
(\ref{quatern5}), exploiting the resolution of the identity (\ref{quatern6}). We identify the domain of the
variables $(z, \overline{z}, \widehat{n})$, appearing in ${\mathfrak Z}(z, \overline{z}, \widehat{n} )$, with
${\mathcal D}\times S^2$ and on it define the measure $d\nu (z, \overline{z} , \widehat{n}) =
1/4\pi\; d\lambda (r)\; d\xi\; d\phi\; \sin\theta\; d\theta$.  Consider the map,
$W: \mathbb C^2 \otimes \h \longrightarrow \mathbb C^2 \otimes L^2 ({\mathcal D}\times S^2 , d\nu )$,
\be
  (W\bPsi )_i (z, \overline{z}, \widehat{n}) = {\mathcal N}(r)^{\frac 12}
    \langle \mathfrak Z (z, \overline{z}, \widehat{n}), i\vert\bPsi\rangle\; , \qquad i = 1,2\; .
\label{quatisom}
\en
Here $\bPsi \in \mathbb C^2 \otimes \h$ is a vector of the form,
$\bPsi = \sum_{\ell = 0}^2\chi^\ell \psi_\ell\;,$ with $\psi_1 , \psi_2 \in \h$.
In view of (\ref{quatern6}), the above map is an isometric embedding of the
Hilbert space $\mathbb C^2 \otimes \h$ onto
a closed subspace of $\mathbb C^2 \otimes L^2 ({\mathcal D}\times S^2 , d\nu )$. We denote this subspace by
$\h_{\text{quat}}$ and elements in it by  $\bF = \sum_{i = 0}^2\chi^i \textsf{F}_i$. Then,
\bea
  \textsf{F}_i (z, \overline{z}, \widehat{n}) & = & \langle\chi^i \vert \bF (z, \overline{z},
  \widehat{n}\rangle_{\mathbb C^2}
        = \langle \mathfrak Z (z, \overline{z}, \widehat{n} ), i\vert\bPsi\rangle_{\mathbb C^2 \otimes \h}
                                           \nonumber\\
          & = & \sum_{\ell = 0}^2\;\sum_{k=0}^\infty{\chi^i}^\dagger u(\widehat{n})\begin{pmatrix}
                 \frac {\overline{z}^k}{\sqrt{x_k !}} & 0 \\ 0 & \frac {z^k}{\sqrt{x_k !}} \end{pmatrix}
                 u(\widehat{n})^* \;\chi^{\ell}\;\langle\phi_k\vert\psi_{\ell}\rangle\;,
\label{quatisom2}
\ena
where we have introduced the matrix,
\be
  u(\widehat{n}) = u_\phi u_\theta =
      \begin{pmatrix} e^{i\frac {\phi}2}\cos{\frac {\theta}2} & ie^{i\frac {\phi}2}\sin{\frac {\theta}2}\\
             ie^{-i\frac {\phi}2}\sin{\frac {\theta}2} & e^{-i\frac {\phi}2}\cos{\frac {\theta}2}\end{pmatrix}\; .
\label{n-matrix}
\en
Next let us introduce the two projection operators on $\mathbb C^2$:
\bea
   \mathbb P_1 (\widehat{n}) & = & u(\widehat{n})\begin{pmatrix} 1 & 0\\0& 0 \end{pmatrix}u(\widehat{n})^*
            =\begin{pmatrix} \cos^2\frac {\theta}2 & -ie^{i\phi}\sin\frac {\theta}2\; \cos\frac {\theta}2\\
              ie^{-i\phi}\sin\frac {\theta}2\; \cos\frac {\theta}2 & \sin^2\frac {\theta}2\end{pmatrix}\nonumber\\
  \mathbb P_2 (\widehat{n}) & = & \mathbb I_2 - \mathbb P_1 (\widehat{n})\; ,
\label{2-projops}
\ena
and the holomorphic functions $f_\ell (z)$, along with their anti-holomorphic counterparts $f_\ell (\overline{z})$,
\bea
    f_\ell (z) & = & \sum_{k=0}^\infty\frac {z^k}{\sqrt{x_k !}}\;\langle\phi_k\vert\psi_\ell\rangle_\h ,
    \qquad \mathbf f (z) = \sum_{\ell = 1}^2\chi^\ell f_\ell (z)\; , \nonumber\\
    f_\ell (\overline{z}) & = & \sum_{k=0}^\infty
        \frac {\overline{z}^k}{\sqrt{x_k !}}\;\langle\phi_k\vert\psi_\ell\rangle_\h\; , \qquad
        \mathbf f (\overline{z}) = \sum_{\ell = 1}^2\chi^\ell f_\ell (\overline{z}) \; .
\ena
Then, it is easy to see that (\ref{quatisom2}) yields the expression,
\be
   \bF (z, \overline{z}, \widehat{n}) = \mathbb P_1 (\widehat{n})\mathbf f (\overline{z})  +
                      \mathbb P_2 (\widehat{n})\mathbf f (z)\; .
\label{holfunct}
\en
Thus, for fixed $\widehat{n}$, the component function $F_i (z, \overline{z}, \widehat{n})$ is a linear combination
of two holomorphic functions $f_1 (z), f_2 (z)$ and their antiholomorphic counterparts.

   Finally, we might note that the reproducing kernel (\ref{matrepker})
in this case is a $2\times 2$ matrix-valued kernel:
\be
  \bK (\overline{z}', z',  \widehat{n}'\; ; \; z, \overline{z}, \widehat{n} ) =
    \sum_{k=0}^\infty \frac 1{x_k !}\;[{\mathfrak Z}(z', \overline{z}', \widehat{n}' )^*]^k\;
    {\mathfrak Z}(z, \overline{z}, \widehat{n} )^k\; ,
\label{quatker1}
\en
with matrix elements,
\be
   \bK (\overline{z}', z',  \widehat{n}'\; ; \; z, \overline{z}, \widehat{n} )_{ij}  =
     [{\mathcal N}(r'){\mathcal N}(r)]^{\frac 12}\;\langle{\mathfrak Z}(z', \overline{z}', \widehat{n}' ); i\; \vert
        \; {\mathfrak Z}(z, \overline{z}, \widehat{n} ); j \rangle\; ,
\label{quatker2}
\en
and satisfying
\begin{align}
   \int_{{\mathcal D}\times S^2} d\nu (z', \overline{z}', \widehat{n}')\;
        \bK (\overline{z}, z,  \widehat{n}\; ; \; z', \overline{z}', \widehat{n}' )\; &
         \bK (\overline{z}', z',  \widehat{n}'\; ; \; z'', \overline{z}'', \widehat{n}'' )\nonumber \\ =
          \bK (\overline{z}, z,  \widehat{n}\; ; \; z'', \overline{z}'', \widehat{n}'' )\; .
\label{quatker3}
\end{align}
Also, in this case, the matrix $\bK (\overline{z}, z,  \widehat{n}\; ; \; z, \overline{z}, \widehat{n} )$
is strictly positive definite for each $z,\;  \overline{z}$, and $\widehat{n}$.

\section{Examples using matrix domains}\label{sec:mat-dom}
\setcounter{equation}{0}

Our last set of examples involve some  matrix domains, which parallel and in some cases include
the results of Section \ref{sec:ex-cliff-alg} as well. As the first example of this type,
let~$\mathcal O_n$ be the unit ball (with respect to the operator
norm) of the space of all $n\times n$ complex matrices:
$$ \mathcal O_n =\{\mathfrak Z\in\mathbb C^{n\times n}\; \vert\; \mathbb I_n - \mathfrak Z\mathfrak Z^*\;\;
\text{ is positive definite} \}.  $$
Let~$\nu$ be a finite measure on $\mathcal O_n$ such that
\begin{equation}
 d\nu(\epsilon \mathfrak Z) = d\nu(\mathfrak Z)\; ,  \qquad\forall \epsilon\in U(1),
\label{tag8}  \end{equation}
and
\begin{equation}
 d\nu(V\mathfrak Z V^*) = d\nu(\mathfrak Z)\; ,  \qquad\forall V \in U(n).
\label{tag9}  \end{equation}
(For~example, one may take $d\nu(\mathfrak Z)=\det[I-\mathfrak Z^* \mathfrak Z]^\alpha\,d\mathfrak Z$, where
$\alpha\ge0$ and $d\mathfrak Z$ is the Lebesgue measure on~$\mathbb C^{n\times n}$, or~any
other measure depending only on the singular values of~$\mathfrak Z$ (see~below).) Let
\begin{equation}
 X_{k\ell} := \int_{\mathcal O_n}  \mathfrak Z^k \mathfrak Z^{*\ell}\, d\nu(\mathfrak Z).
\label{tag10}  \end{equation}
Then by~(\ref{tag8})
$$ X_{k\ell} = \into  (\epsilon \mathfrak Z)^k (\epsilon \mathfrak Z)^{*\ell}\,d\nu(\mathfrak Z)
= \epsilon^{k-\ell} X_{k\ell}  $$
for all $\epsilon\in U(1)$, implying that $X_{k\ell}=0$ if $k\neq \ell$. Furthermore,
by~(\ref{tag9})
\begin{align*}
X_{kk} &= \into  (V\mathfrak ZV^*)^k (V\mathfrak ZV^*)^k\, d\nu(\mathfrak Z) \\
&= \into V \mathfrak Z^k \mathfrak Z^{* k} V^* \, d\nu(\mathfrak Z) \\
&= V X_{kk} V^*, \end{align*}
so~$[X_{kk},V]=0$ for all $V\in U(k)$. This implies that $X_{kk}=q_k \mathbb I_n$ for
some $q_k\in\mathbb C$. Plainly $q_k>0$, since the integrand in (\ref{tag10}) is positive
definite for $k=\ell$. Thus we can take
\begin{equation}
 F_k(\mathfrak Z) := q_k^{-1/2} \mathfrak Z^k = \frac {\mathfrak Z^k}{\sqrt{x_k!}}\;,  \qquad
 x_k = \frac {q_k}{q_{k-1}}\;,
 \label{mat-dom-F}
 \end{equation}
with the assumption that $d\nu$ has been normalized so that $q_1 = 1$.
The normalization condition (\ref{F-cond1}) takes the form
\begin{equation}
  \NN (\mathfrak Z) = \sum_k \frac {\Tr[\mathfrak Z^{* k} \mathfrak Z^k]}{x_k !} <\infty .
\label{tag11}
\end{equation}
We~claim that this holds, for all $\mathfrak Z\in\mathcal O_n$, as~soon as the support of $\nu$
is all of~$\mathcal O_n$. To~see this, recall that any $n\times n$ matrix $\mathfrak Z$ can be
written in the form
\begin{equation}
 \mathfrak Z= V \cdot \operatorname{diag}(r_1,\dots,r_n)\cdot W^*,  \label{tag12}
\end{equation}
where $V,W\in U(n)$ and $0\le r_n\le\dots\le r_1=\|\mathfrak Z\|$ are the singular
numbers of~$\mathfrak Z$ (i.e.~eigenvalues of~$\mathfrak Z \mathfrak Z^*$); the~unitary matrices $V,W$ need
not be uniquely determined by~$\mathfrak Z$ (they are iff all the $r_j$ are different),
but~the diagonal part~is. We~then have
$$ r_n \|\bv\| \le \|\mathfrak Z \bv\| \le r_1 \|\bv\|\; ,  \qquad \forall \; \bv\in\mathbb C^n\; .  $$
Taking $\bv$ to be a unit vector, it follows that
\begin{equation}
 q_k = \spr{\bv\; \vert\;  X_{kk}\bv}_{\mathbb C^n} = \into \|\mathfrak Z^k \bv\|^2 \,d\nu(\mathfrak Z)
\label{mat-VCS-coeff}
\end{equation}
satisfies
$$ \into r_n(\mathfrak Z)^{2k} \,d\nu(\mathfrak Z) \le q_k \le
\into r_1(\mathfrak Z)^{2k} \,d\nu(\mathfrak Z).  $$
Taking $k$-th roots and using the fact that $\|f\|_{L^k(d\nu)}\to
\|f\|_{L^\infty(d\nu)}$ for any finite measure~$\nu$, we~see that
$$ \|r_n\|^2_{L^\infty(d\nu)} \le \liminf_{k\to\infty} q_k^{1/k}
\le \limsup_{k\to\infty} q_k^{1/k} \le  \|r_1\|^2_{L^\infty(d\nu)}.  $$
Thus if $\operatorname{supp}\nu=\mathcal O_n$, then $\lim_{k\to\infty} q_k^{1/k}=1$.
Since $\Tr[\mathfrak Z^{* k}\mathfrak Z^k]\le n\|\mathfrak Z^{* k}
\mathfrak Z^k\| \le n\|\mathfrak Z\|^{2k}$, it~follows that the
series (\ref{tag11}) converges $\forall \;\mathfrak Z\in\mathcal O_n$.

  Explicitly, for the matrix domain $\mathcal O_n$ we then have the  VCS,
\begin{equation}
\vert \mathfrak Z ; i\rangle = \NN (\mathfrak Z )^{-1}\sum_{k=0}^\infty\frac {\mathfrak Z^k}{\sqrt{x_k!}}
   \chi^i\otimes\phi_k\; , \qquad \mathfrak Z \in \mathcal O_n\; ,
\label{mat-dom-VCS}
\end{equation}
where $\NN(\mathfrak Z )$ is given by (\ref{tag11}),  the $\chi^i$ form an orthonormal
basis in $\mathbb C^n$ and $x_k$ is given via (\ref{mat-dom-F})
and (\ref{mat-VCS-coeff}). The reproducing kernel,
\begin{equation}
\bK (\mathfrak Z^* ; {\mathfrak Z}^\prime) =
  \sum_{k=0}^\infty\frac {[\mathfrak Z^*]^k\; [\mathfrak Z^\prime]^k}{\sqrt{x_k!}}\; ,
\label{mat-dom-ker}
\end{equation}
is an $n\times n$ matrix kernel, with $\bK (\mathfrak Z^*  ; \mathfrak Z ) > 0$, for all $\mathfrak Z$.

For~measures $\nu$ for which (\ref{tag11}) fails, one~can again save the situation by
the same trick as in Section \ref{sec:ex-planch-isom}: namely, fix~some measure space $(R,dr)$,
consider $X = R\times\mathcal O_n$, and set
$$ F_k(x) = f_k(r) q_k^{-1/2} \mathfrak Z^k, \qquad x=(r,\mathfrak Z)\in X,  $$
with some fixed unit vectors $f_k\in L^2(R,dr)$. Then once again
$$ \iint_{R\times\mathcal O_n}F_k(x) F_\ell(x)^*  \, dr \,d\nu(\mathfrak Z) = \delta_{k\ell}\; \mathbb I_n,   $$
provided the $f_k$ are chosen so that the condition (\ref{tag11})
$$\NN (x) =  \sum_k |f_k(r)|^2 q_k^{-1} \Tr[\mathfrak Z^{* k}\mathfrak Z^k]
< \infty\; ,  \qquad \forall (r,\mathfrak Z ) \in X\; ,   $$
is satisfied. This can always be achieved, no~matter what $q_k$ and
$\Tr[\mathfrak Z^{* k}\mathfrak Z^k]$ are.

%\medskip
\newpage

{\sl Remarks:}
\begin{enumerate}

\item  The~last example can also be generalized to any domain
$\mathcal O \subset\mathbb C^{n\times n}$ which is invariant under the transformations
$\mathfrak Z\mapsto V\mathfrak Z W^*$,\; $\forall\; V,W\in U(n)$, and any measure $\nu$ on $\mathcal O$
satisfying (\ref{tag8}) and~(\ref{tag9}) and such that $\int_{\mathcal O}
\|\mathfrak Z^{*\ell}\mathfrak Z^k\|\,d\nu(\mathfrak Z)$ is finite
$\forall \; (k,\ell)$. The~condition (\ref{tag11}) is  satisfied whenever
supp$\,\nu=\mathcal O$; otherwise one again needs to introduce the auxiliary
measure space~$R$.

\item We~can also deal in the same way with the case when $\mathcal O$ is the
unit ball of $n\times n$ complex {\em symmetric} or {\em anti-symmetric}
matrices,~i.e. one of the domains
\begin{align*}
\mathcal O_n^{\;\text{sym}} &:=
\{\mathfrak Z\in\mathbb C^{n\times n}\; \vert \; \|\mathfrak Z\|<1 \text{ and } \mathfrak Z^T =\mathfrak Z \} ,  \\
\mathcal O_n^{\;\text{a-sym}} &:=
\{\mathfrak Z\in\mathbb C^{n\times n}\; \vert\; \|\mathfrak Z\|<1 \text{ and } \mathfrak Z^T =-\mathfrak Z \} .
\end{align*}
In~this case, (\ref{tag9}) should be required to hold only for all symmetric
unitary matrices~$V$; then the argument after (\ref{tag10}) implies that
$[X_{kk},V]=0$ for all such matrices, which is still sufficient for concluding
that $X_{kk}$ is a multiple of the identity since $X_{kk}$ must now also be a
symmetric matrix.

\item  Observe that if we require, instead of~(\ref{tag9}), that  $d\nu(V\mathfrak Z W^*)=
d\nu(\mathfrak Z)$, \;  $\forall\; V,W\in U(n)$, then it follows from (\ref{tag12}) that $d\nu$  admits the
measure disintegration,
$$ d\nu(\mathfrak Z) = d\mu(r_1,\dots,r_n) \, d\Omega_n (V) \, d\Omega_n (W)  $$
(with $\mathfrak Z$ decomposed as in (\ref{tag12})), where  $d\Omega_n$ is the Haar measures on~$U(n)$
and $d\mu$ some measure on
$\mathbb R^n$ invariant under permutations of the coordinates. This is reminiscent of
the ``polar decomposition'' (\ref{cliff-vcs2}).

\end{enumerate}

  We propose to report on these cases in a future publication. However, as one last interesting
example, consider the set, $\mathcal O^{\;\text{nor}}_n$ of all $n\times n$ complex, normal matrices, i.e.,
matrices $\mathfrak Z$ satisfying $\mathfrak Z^* \mathfrak Z = \mathfrak Z \mathfrak Z^*$. Such a
matrix has the decomposition,
\be
 \mathfrak Z = V\cdot \text{diag}(r_1 e^{i\theta_1},r_2 e^{i\theta_2},\ldots , r_n e^{i\theta_n})\cdot V^*\; ,
    \qquad  V \in U(n), \;\; r_i \geq 0, \;\; 0\leq \theta_i < 2\pi\;  .
\label{nor-mat-dec}
\end{equation}
Let $\lambda_i , \; i=1,2, \ldots , n$, be a set of positive measures on on $\mathbb R^+$, satisfying the moment
problems,
\be
 \int_0^{L_i}d\lambda_i (r)\; r^{2k} = \frac {x^i_k !}{2\pi}\; ,\qquad i =1,2, \ldots , n\; ,
\label{mom-eqs}
\end{equation}
where, for fixed $i , \;\;   x^i_k ! = x^i_1 x^i_2 \ldots x^i_k , \; \;  x^i_0 ! = 1$
and $x^i_1 = 1$. Also, we assume as usual,
that $L_i > 0$ is the radius of convergence of the series $\sum_{k = 0}^\infty\frac {y^k}{\sqrt{x^i_k !}}$.
With $d\Omega_n$ the Haar measure on $U(n)$ (normalized to one), define the measure $d\nu$ and the
domain $\mathcal D$ by
\be
    d\nu (\mathfrak Z ) = \prod_{i=1}^n d\lambda_i (r_i )\; d\theta_i\; d\Omega_n (V)\; , \qquad
 \mathcal D = \prod_{i=1}^n [0, L_i )\times [0, 2\pi )^n \times U(n)\; .
\label{nor-mat-meas}
\end{equation}

  Let us calculate the integral
\be
  X_{k \ell} = \int_{\mathcal D}d\nu (\mathfrak Z )\; \mathfrak Z^k \mathfrak Z^{*\ell}\; .
\label{xkl-integ}
\end{equation}
  We have the  result:
 \belem
 \be
   X_{k \ell} = \frac 1n \sum_{i=1}^n x^i_k !\; \delta_{k\ell}\;\mathbb I_n\; .
 \label{xkl-integ2}
 \end{equation}
 \enlem

 \prf
   Using the decomposition (\ref{nor-mat-dec}), the definition of the measure and domain in
(\ref{nor-mat-meas}) and the moment equations (\ref{mom-eqs}), we see that
\beano
   X_{k \ell} & = & \int_0^{L_1}\!\! d\lambda_1 \int_0^{L_2}\!\! d\lambda_2 \ldots \int_0^{L_n}\!\! d\lambda_n
            \int_0^{2\pi}\!\! d\theta_1 \int_0^{2\pi}\!\! d\theta_2\ldots \int_0^{2\pi}\!\! d\theta_n\\
           &\times & \int_{U(n)}
            \!\!d\Omega_n (V)\; V\cdot\text{diag}(r_1^{k + \ell} e^{i(k-\ell)\theta_1},
            \;r_2^{k + \ell} e^{i(k-\ell)\theta_2},\;
            \ldots , \; r_n^{k + \ell} e^{i(k-\ell)\theta_n})\cdot V^*\\
            & = & \int_{U(n)}d\Omega_n (V)\; V\cdot \text{diag}(x^1_k !,\;  x^2_k ! , \;  \ldots ,
            \;x^n_k !)\cdot V^*\; \delta_{k\ell}\\
            & = & \sum_{i=1}^n x^i_k !\;\int_{U(n)}d\Omega_n (V)\;V \vert e_i\rangle\langle e_i \vert V^*\;
            \delta_{k\ell}\; ,
\enano
where $\{e_i\}_{i=1}^n$ is the canonical orthonormal basis of $\mathbb C^n$:
$$ e_1 = \begin{pmatrix} 1\\ 0\\ \vdots\\ 0\end{pmatrix}\;, \quad
 e_2 = \begin{pmatrix} 0\\ 1\\ \vdots\\ 0\end{pmatrix}\;, \quad \ldots \; , \quad
 e_n = \begin{pmatrix} 0\\ 0\\ \vdots\\ 1\end{pmatrix}\; . $$
From the general orthogonality relations, holding for compact groups (see, for example, \cite{aag_book}),
we know that
$$
  \int_{U(n)}d\Omega_n (V)\;V \vert e_i\rangle\langle e_i \vert V^* = \frac 1n \;\mathbb I_n\; ,
  \qquad \forall i\; , $$
from which (\ref{xkl-integ2}) follows.
\qed

\medskip
\medskip

Setting $F_k (\mathfrak Z ) = \mathfrak Z^k /\sqrt{q_k}$, where now $q_k = x_k ! / n$ (see
(\ref{mat-VCS-coeff})), it is straightforward
to build the associated
VCS. Indeed, we have:
\betheo
  The vectors,
\be
   \vert \mathfrak Z ; i\rangle = \mathcal N (\mathfrak Z )^{-\frac 12}\sum_{k=0}^\infty
         \frac {\mathfrak Z^k}{\sqrt{q_k}}\chi^i\otimes \phi_k \; , \qquad \mathcal N (\mathfrak Z )
         = \sum_{k=0}^\infty \frac {\text{Tr}[\vert \mathfrak Z \vert^{2k}]}{q_k}\; ,
\label{nor-VCS}
\end{equation}
for $\mathfrak Z \in \mathcal D$ and $i=1,2, \ldots , n$, form a family of VCS in $\mathbb C^n\otimes \h$.
\entheo

In particular, if we take
\be
  L_i = \infty\; , \quad d\lambda_i (r) = \frac 1\pi e^{-r^2}r\; dr\; , \qquad i=1,2, \ldots , n\; ,
\label{nor-cancs1}
\end{equation}
then
\be
 \mathcal D = \mathcal O^{\;\text{nor}}_n \simeq \mathbb C^n\times U(n)\;, \quad  q_k = k!  \quad \text{and}
 \quad  \mathcal N (\mathfrak Z) =
             e^{\text{Tr}[\vert\mathfrak Z\vert^2 ]}\;,
\label{nor-cancs2}
\end{equation}
while the measure $d\nu$ becomes,
\be
  d\nu (\mathfrak Z ) = \frac {e^{- \text{Tr}[\vert\mathfrak Z\vert^2 ]}}{(2\pi i)^n}\prod_{j=1}^n
     d\overline{z}_j\wedge dz_j\; d\Omega_n\; , \qquad z_j = r_j e^{i\theta_j}\;,\;\;  j =1, 2, \ldots , n\; .
\label{nor-cancs3}
\end{equation}
The corresponding VCS,
\be
   \vert\mathfrak Z; i\rangle = e^{-\frac 12 \text{Tr}[\vert\mathfrak Z\vert^2]}\sum_{k=0}^\infty
      \frac{\mathfrak Z^k}{\sqrt{k!}}\chi^i\otimes\phi_k\; , \qquad i=1,2,\ldots , n\; .
\label{normatCS}
\end{equation}
are then the analogues of the canonical coherent states (\ref{CCS}) over this domain, which
we now analyze in some detail.

The VCS (\ref{normatCS}) satisfy the resolution of the identity,
\be
 \frac 1{(2\pi i)^n}\sum_{i=1}^n \int_{\mathcal O_n^{\; \text{nor}}}
     \prod_{j=1}^n d\overline{z}_j\wedge dz_j\; d\Omega_n (V)\; e^{- \text{Tr}[\vert\mathfrak Z\vert^2 ]}\;
        \vert\mathfrak Z; i\rangle \langle\mathfrak Z; i\vert = \mathbb I_n \otimes I_\h\; .
\label{normCS-resolid}
\end{equation}
The associated reproducing kernel is
\be
 \bK (\mathfrak Z^* , \mathfrak Z' ) = \sum_{k=0}^\infty\frac {[\mathfrak Z^*]^k [\mathfrak Z']^k}{k!}\; ,
\label{normCSrepker}
\end{equation}
which satisfies $\bK (\mathfrak Z^* , \mathfrak Z ) > 0$, implying that the VCS (\ref{normatCS}),
for fixed $\mathfrak Z$
and $i=1,2, \ldots , n$, are linearly independent. Furthermore,
\begin{align}
 & \frac 1{(2\pi i)^n}\sum_{i=1}^n \int_{\mathcal O_n^{\; \text{nor}}}
     \prod_{j=1}^n d\overline{z''}_j\wedge dz''_j\; d\Omega_n (V'')\; e^{- \text{Tr}[\vert\mathfrak Z''\vert^2 ]}\;
   \bK (\mathfrak Z^* , \mathfrak Z'' ) \;\bK (\mathfrak Z^{\prime\prime *} , \mathfrak Z' )\nonumber \\
   & \qquad \qquad  =
   \bK (\mathfrak Z^* , \mathfrak Z' )\; .
\label{normCSrepker2}
\end{align}

  Introducing next the usual creation and annihilation operators, $a^\dag , \; a$ on $\h$,
$$
   a^\dag \phi_n = \sqrt{n+1}\phi_{n+1}\; , \qquad a\phi_n = \sqrt{n}\phi_{n-1}\; , $$
we note that
\be
    \mathfrak Z^k \chi^i\otimes \phi_k  = \frac {(\mathfrak Z\otimes a^\dag)^k}{\sqrt{k!}}\chi^i\otimes\phi_0 \qquad
\text{and} \qquad e^{\mathfrak Z^* \otimes a}\chi^i\otimes\phi_0 = \chi^i\otimes\phi_0\;.
\label{normCS2}
\end{equation}
Furthermore, since
\be
  [\mathfrak Z^*\otimes a , \; \mathfrak Z\otimes a^\dag ] = V\cdot\text{diag}(r_1^2 ,
   r_2^2 , \cdots ,  r_n^2 )\cdot V^*\otimes I_\h\; ,
\label{normCS3}
\end{equation}
and since both $\mathfrak Z$ and $\mathfrak Z^*$ commute with $V\cdot\text{diag}(r_1^2 ,
r_2^2 , \cdots , r_n^2 )\cdot V^*$ (this is clear from the form of $\mathfrak Z$ given in
(\ref{nor-mat-dec})), we may use the well-known Baker-Campbell-Hausdorff
identity,
$$ e^{A+B} = e^{-\frac 12 [A,B]}e^A e^B\;, $$
which holds when both $A$ and $B$ commute with $[A,B]$, to get
\be
  \mathbb D (\mathfrak Z):=  e^{\mathfrak Z\otimes a^\dag - \mathfrak Z^*\otimes a} =
  e^{-\frac 12 V\cdot\;\text{diag}(r_1^2 ,\;
   r_2^2 ,\; \cdots ,\; r_n^2 )\cdot\; V^*}\;e^{\mathfrak Z\otimes a^\dag}  e^{- \mathfrak Z^*\otimes a}\; .
\label{normCS4}
\end{equation}
Combining (\ref{normatCS}), (\ref{normCS2}) and (\ref{normCS4}), we finally obtain
\be
  \vert \mathfrak Z ; i\rangle = e^{-\frac 12 VT V^*}\; \mathbb D (\mathfrak Z )
  \chi^i\otimes \phi_0\; , \qquad
  \mathfrak Z \in \mathcal O^{\; \text{nor}}_n\; ,
\label{normCS5}
\end{equation}
where $T$ is the diagonal matrix,
\begin{equation}
T = \text{diag}(a_1, a_2 , \ldots , a_n)\; , \qquad
  a_i = \sum_{j=1}^n r_j^2 - r_i^2\; .
\label{normCS6}
\end{equation}
The operator  $\mathbb D (\mathfrak Z)$ is unitary on $\mathbb C^n\otimes\h$ and may also be written in
the suggestive form,
\be
 \mathbb D (\mathfrak Z) =V\cdot \text{diag}(D(z_1 ), \; D(z_2 ), \; \ldots , \; D(z_n))\cdot V^*\;, \qquad
   z_j = r_j e^{i\theta_j}\; ,
\label{normdispop}
\end{equation}
where $D(z) = e^{za^\dag - \overline{z}a}\;,\;\; z \in \mathbb C$, is the so-called  {\em displacement
operator}, defined on $\h$. By analogy we shall refer to $\mathbb D (\mathfrak Z)$ as the {\em matrix
displacement operator\/}. Since the $D(z), \; z \in \mathbb C$, realize a unitary projective
representation of the Weyl-Heisenberg
group, for each fixed $V \in U(n)$, the operators $\mathbb D (\mathfrak Z)$ realize an $n$-fold reducible
projective representation of this group on $\mathbb C^n \otimes\h$. Equation (\ref{normCS5}) is the analogue
of the relation $\vert z \rangle = D(z)\phi_0$, which holds for the canonical coherent states (\ref{CCS}).

  The analysis of Section \ref{sec-analyt-prop} can also be repeated here almost verbatim. The map
$W: \mathbb C^n\otimes \h \longrightarrow \mathbb C^n\otimes L^2 (\mathcal O^{\; \text{nor}}_n , d\nu )$,
where,
\be
 (W\bPsi)(\mathfrak Z^* )_i = e^{\frac 12 \text{Tr}[\vert\mathfrak Z\vert^2]}\;\langle \mathfrak Z; i\;\vert\; \bPsi \rangle\; ,
 \qquad i=1,2, \ldots , n\; ,
\label{nor-mat-isom}
\end{equation}
is an isometric embedding of $\mathbb C^n\otimes\h$ onto a (closed) subspace of
$\mathbb C^n\otimes L^2 (\mathcal O^{\; \text{nor}}_n , d\nu )$, which we denote by $\h_\text{nor}$. To
study the nature of this subspace, let us write elements in it as $\bF = \sum_{\ell =1}^n \chi^\ell \textsf{F}_\ell\;,$
with $\textsf{F}_\ell \in L^2 (\mathcal O^{\; \text{nor}}_n , d\nu )$. Setting $\bPsi =
\sum_{\ell =1}^n \chi^\ell\psi_\ell\;,$ $\psi_\ell \in \h\;,$ and $\bF = W\bPsi\;,$ we have,
$$
  \textsf{F}_i (\mathfrak Z^* ) = \mathcal N (\mathfrak Z)^{\frac 12}\;
    \langle \mathfrak Z ; i \; \vert\; \bPsi \rangle\; . $$
For $\ell =1,2, \ldots , n$, let $f_\ell$ denote the analytic function,
\be
  f_\ell (z) = \sum_{k=0}^\infty\frac {\langle\phi_k \vert\psi_\ell\rangle}{\sqrt{k!}}z^k\;,
         \qquad z \in \mathbb C\; ,
\label{normanalfcn}
\end{equation}
and $\mathbf f = \sum_{\ell = 1}^n \chi^\ell f_\ell$. Let $\mathbb P_j (V)$ be the one-dimensional projection
operator (on $\mathbb C^n$),
\be
    \mathbb P_j (V) = V\vert\chi^j\rangle\langle \chi^j\vert\; V^*\; , \qquad j =1,2, \ldots , n\; .
\label{normproj}
\end{equation}
Then once again one can show that,
\be
  \bF (\mathfrak Z^* ) = \sum_{j=1}^n \mathbb P_j (V)\mathbf f (\overline{z}_j)\; ,
\label{normanalfcn2}
\end{equation}
where the $z_j = r_j e^{i\theta_j}$ are the variables appearing in the decomposition of the matrix $\mathfrak Z$ in
(\ref{nor-mat-dec}). The above relation should be compared to (\ref{holfunct}). Thus, $\h_\text{nor}$
consists of linear combinations of anti-analytic functions
in the variables $z_j$ and square-integrable with respect to the measure $d\nu$ in (\ref{nor-cancs3}).
It is now abundantly clear that all these results reduce to their well-known
counterparts  for the canonical coherent states (\ref{CCS}) when $n =1$.

\section*{Acknowledgements}
   The authors would like to thank Marco Bertola for useful discussions.
The first named author (STA) benefitted from a Natural Sciences and Engineering Research Council
of Canada grant, while the second author (ME) was supported by a GA {\v C}R grant (No.~201/03/0041).

   %%%%%%%%%%%%%%%%%%%%%%%%%%%%%%%%%%%%%%%%%%%%%%%%%%%%%%

\end{document}